\pdfoutput=1 
\documentclass[a4paper,conference]{IEEEtran}

\newif\iffull
\fulltrue
\IEEEsettopmargin{t}{30mm}
\IEEEquantizetextheight{c}
\IEEEsettextwidth{14mm}{14mm}
\IEEEsetsidemargin{c}{0mm}

\usepackage[utf8]{inputenc}
\usepackage[T1]{fontenc}
\usepackage{url}
\usepackage{ifthen}
\usepackage{cite}
\usepackage[cmex10]{amsmath}

\usepackage{amsthm}
\usepackage{algorithmic, algorithm}
\usepackage[english]{babel}
\usepackage[table]{xcolor}
\usepackage{enumitem}
\usepackage{amssymb}
\usepackage{xcolor}
\usepackage{graphicx}
\usepackage{tikz}
\usepackage[hidelinks]{hyperref}
\usepackage[font=normalsize]{caption}

\theoremstyle{definition}

\newtheorem{lemma}{Lemma}
\newtheorem{definition}{Definition}
\newtheorem{corollary}{Corollary}

\renewcommand{\v}[1]{\mathbf{#1}}

\newcommand{\full}[1]{\iffull #1 \fi}
\newcommand{\short}[1]{\iffull \else #1 \fi}

\title{\huge M-DAB: An Input-Distribution Optimization Algorithm for Composite DNA Storage by the Multinomial Channel}

\author{\IEEEauthorblockN{Adir Kobovich, Eitan Yaakobi and Nir Weinberger}
\IEEEauthorblockA{\textit{Technion -- Israel Institute of Technology, Haifa, Israel} \\
Email: adir.k@campus.technion.ac.il, yaakobi@cs.technion.ac.il, nirwein@technion.ac.il}
}

\IEEEoverridecommandlockouts
\IEEEaftertitletext{\vspace{-5pt}}

\begin{document}
\maketitle

\begin{abstract}
Recent experiments have shown that the capacity of DNA storage systems may be significantly increased by 
synthesizing composite DNA letters. In this work, we model a DNA storage channel with composite inputs as a \textit{multinomial channel}, and propose an optimization algorithm for its capacity achieving input distribution, for an arbitrary number of output reads. The algorithm is termed multidimensional dynamic assignment Blahut-Arimoto (M-DAB), and is a generalized version of the DAB algorithm, proposed by Wesel et al. \cite{wesel2018efficient} developed for the binomial channel. We also empirically observe a scaling law behavior of the capacity as a function of the support size of the capacity-achieving input distribution.

\end{abstract}

\section{Introduction}
\label{sec:introduction}
DNA storage \cite{church2012next, goldman2013towards} is an emerging technology that involves converting digital information into nucleotide sequences with quaternary encoding, represented by the letters $A$, $C$, $G$, and $T$. The sequence, known as a \textit{strand}, is written via \textit{synthesis} and is read via \textit{sequencing}. An intriguing aspect of this process is the generation of multiple copies of the same strand during the synthesis process. In this paper, we focus on one approach to harnessing this redundancy, achieved by introducing the concept of \textit{composite DNA letters} \cite{anavy2019data, choi2019high, preuss2021data, zhang2022limited, yan2023scaling}. These composite letters consist of different mixtures of nucleotides, and have been successfully utilized in data encoding experiments \cite{anavy2019data, choi2019high, yan2023scaling}. In theory, using composite DNA letters can dramatically increase the capacity of the DNA storage channel, since while the capacity of simple $4$-letter DNA encoding is bounded by $\log(4)=2$ bits per channel use, the capacity of composite DNA encoding is unbounded.  Furthermore, the larger capacity enables encoding data in shorter strands, which is particularly effective in DNA storage, due to the high cost of the synthesis process \cite{choi2019high}, and the nature of the process, in which the error probability increases as the strand gets longer \cite{bornholt2017toward}. 

The process of writing a composite letter and randomly reading $n$ copies can be modeled as the operation of a noisy channel. In this channel, the input is a probability vector of length $k=4$ (in the case of DNA letters), which represents a mixture of nucleotides. The channel output is distributed as a multinomial random variable, with $n$ event trials and event probabilities given by the input vector. The input represents the expected frequency of occurrences of each of the four nucleotides in the $n$ output copies. We thus refer to this channel as the \textit{multinomial channel}. The maximal storage rate of information possible over this channel is the \textit{capacity} of the channel. As is well-known \cite{cover1999elements}, the capacity is obtained by maximizing the mutual information between the input and output of the channel, over all feasible choices of input distributions, that is, distributions over the $(k-1)$-dimensional probability simplex. 

In this paper, we consider the problem of determining the capacity-achieving input distribution (CAID) when the number of reads $n$ is finite. Since the output alphabet of this channel is discrete, there exists a CAID with support of finite cardinality. Thus, the CAID can be parameterized by a finite set of points in the $(k-1)$-dimensional simplex (\textit{locations}) and their respective probabilities (\textit{weights}). If one then further fixes the locations, then the channel is reduced to a discrete memoryless channel (DMC), and the optimal probabilities can be computed by the Blahut-Arimoto algorithm \cite{blahut1972computation}. Consequently, the main computational challenge is to identify the optimal locations in the $(k-1)$-dimensional simplex.

Previous works have addressed this challenge for the special case of $k=2$, known as the \textit{binomial channel} \cite{komninakis2001capacity}, and for a variation of it called the \textit{particle-intensity channel} \cite{farsad2020capacities}. These works introduce an algorithm for finding the CAID, termed the \textit{Dynamic Assignment Blahut-Arimoto} (DAB) algorithm \cite{wesel2018efficient}. In this paper, we propose a generalization of the DAB algorithm, which finds the CAID for the multinomial channel ($k>2$). We focus on the challenges associated with the multidimensionality of the channel symbols, and refer our algorithm as the \textit{multidimensional dynamic assignment Blahut-Arimoto} (M-DAB).
Using M-DAB, we compute the CAID for various values of $n$. These CAIDs can directly be used in coded DNA storage systems to obtain improved coding rates.  In addition, we evaluate the cardinality of the support of the CAID as a function of the mutual information. We show that this cardinality matches the scaling law behavior, recently conjectured in \cite{abbott2019scaling}.

This paper is organized as follows. Section~\ref{sec:Problem Definition} defines the multinomial channel and the optimization problem. Section~\ref{sec:CAID} studies properties of the multinomial channel CAID, which be later used in our algorithm. Section~\ref{sec:MultinomialDAB} introduces the DAB algorithm as preliminary work and presents the M-DAB algorithm including the challenges associated with multidimensional inputs. Section~\ref{sec:results} shows the achieved CAID and corresponding channel capacities. Lastly, in Section~\ref{sec:conclusion} we discuss open problems. \short{Due to limited space, full proofs can be found in the extended version~\cite{Extended}.}

\section{Multinomial Channel Problem Definition}
\label{sec:Problem Definition}

In this section, we formally define the multinomial channel and the resulting CAID optimization problem. The multinomial channel \textit{input} alphabet is the $(k-1)$-dimensional probability simplex, given by
$\Delta_{k}:=\{x\in\mathbb{R}_{+}^{k}\mid\sum\limits _{i=1}^{k}x_{i}=1\}$. The multinomial channel \textit{$n$-output} alphabet is the set of all multisets with cardinality $n$ over $[k]$, given by $\mathcal{Y}_{n,k}:=\{y\in\mathbb{Z}_{+}^{k}\mid\sum\limits _{i=1}^{k}y_{i}=n\}$, whose cardinality satisfies $|\mathcal{Y}_{n,k}| = \binom{n+k-1}{k-1}$. For a given input $x\in\Delta_k$, the output of the \textit{multinomial} channel is given by $Y\sim \mathrm{Multinomial}(n,x)$, that is, given input $x \in \Delta_k$ the output $y \in \mathcal{Y}_{n,k}$ obeys the following transition probability 
\begin{equation*} \label{eq: multinomial channel}
 P_{Y|X}^{(n,k)}(y|x)= \frac{n!}{\prod_{j = 1}^{k} y_j!} \prod_{j = 1}^{k}x_j^{y_j}.
\end{equation*}
The channel is referred to as \textit{binomial} if $k=2$.
Hence, if the input is a composite letter $x\in\Delta_k$, then the expected number of times that the $i$-th alphabet letter appears in the output strand is $nx_i$.
We remark that this channel only models the randomness in the output due to sampling of the input, but does not model additional noise in the reading process. If the reading process can be modeled as a symmetric DMC, with a total flip probability of $\epsilon$ (thus $\frac{\epsilon}{k-1}$ to each of the other $k-1$ letters), then the result is simply a  $\mathrm{Multinomial}(n,x \ast \epsilon)$ channel, where
\begin{equation*}
   (x \ast \epsilon)_i := x_i(1-\epsilon)+\epsilon(1-x_i) \hspace{3mm} \text{for all} \hspace{3mm} i\in [k]. 
\end{equation*}
It is thus straightforward to extend our algorithm and results to this case too. 
For convenience, we henceforth consider the noiseless channel.

Our main objective is to find the CAID of the multinomial channel, i.e., to find the input distribution that maximizes the mutual information. 
Let $\mathcal{F}_k$ be the set of all input distributions supported on the input alphabet $\Delta_k$. Specifically, we aim to solve the following optimization problem of the CAID of the  multinomial channel: 
\begin{equation}
C_{n,k} := \max\limits_{f_X\in\mathcal{F}_k} I(X;Y).
\label{eq: capacity optimization problem}
\end{equation}
Alternatively, the dual problem to \eqref{eq: capacity optimization problem}, also known as the Csisz\'{a}r minimax capacity theorem \cite{csiszar2011information}, is given by 
\begin{equation} \label{eq: dual problem alt}
   C_{n,k} = \min\limits_{P_{Y}} \max\limits_{x\in\Delta_k} D(P_{Y|X=x}||P_{Y}),
\end{equation}
where $D(\cdot||\cdot)$ denotes the Kullback–Leibler (KL) divergence, and $P_Y$ is a distribution over $\mathcal{Y}_{n,k}$. 

\section{Properties of the Capacity-Achieving Input Distribution  }
\label{sec:CAID}
For any finite $n$, there is no analytical solution for the capacity and CAID of the multinomial channel. Thus,  we next derive a few properties of this CAID, which will be useful later on, e.g., in reducing the size of the optimization input space. Interestingly, the same capacity arises in universal coding \cite{davisson1973universal}, where it has been demonstrated that the CAID is asymptotically proportional to Jeffrey's prior \cite{clarke1994jeffreys}, and the following asymptotic expression holds \cite{xie1997minimax}
\begin{equation}
\lim_{n \to \infty} \left( C_{n,k} - \frac{k-1}{2} \log \left(\frac{n}{2\pi e} \right) \right) = \log \left(\ \frac{\Gamma^k (1/2)}{\Gamma (k/2)} \right) ,
\end{equation}
where $\Gamma(z) := \int_{0}^{\infty} e^{-t}t^{z-1} dt$ is the Gamma function. 

The first property shows that the CAID can be atomic with finite support. We modify the result of \cite{witsenhausen1980some}, which was proven using Dubins' theorem \cite{dubins1962extreme}: 
\begin{lemma}
Consider a channel, with an input $X$ taking values in $\Delta_k$ for some  $k>1$, and a discrete finite output alphabet $Y$. Assume the transition probability distribution function $x\to P_{Y|X}(y|x)$ is continuous for each $y\in Y$.
Then, there exists a CAID supported on less than $|Y|$ points in $\Delta_k$.
\label{lemma:DMC}
\end{lemma}
\full{
\begin{IEEEproof} 
    While the original proof in \cite{witsenhausen1980some} considered continuous input $X$ taking values in $[0,1]$, which is equal to the $\Delta_2$, all the claims are holding to any $\Delta_k$ as it is a finite dimensional convex and compact set. 
\end{IEEEproof}
 }
\begin{corollary}
There is a CAID with finite support size $m \leq |\mathcal{Y}(k,n)|$. The corresponding input distribution is given by
\begin{equation} \label{eq: atomic input distribution} 
\mathnormal{f}^*_X(x)=\sum\limits_{i = 1}^{m} p^*_{i} \delta(x-x^{(i)}),
\end{equation}
and $\delta(x)$ is the Dirac delta function. That is, ${f}^*_X(x)$ is an atomic distribution. 
\label{cor:DMC}
\end{corollary}

The second property pertains to the set of input symbols at the vertices of the simplex. In the context of DNA encoding, the vertices symbols are the non-composite letters. 

\begin{lemma} 
$C_{n,1}$ is achieved by a uniform distribution on the $k$ vertices of $\Delta_k$. 
\label{lemma:init}
\end{lemma} 

\full{
\begin{IEEEproof}
Notice $C_{n=1,k}$ can be viewed as a single use of a channel with input alphabet with size $k$. Such channel capacity is bounded from above with $\log k$. One readily show that uniform distribution on all the $k$ vertices achieves the upper bound.
\end{IEEEproof}
}

This property allows us to use the CAID for $n=1$ as an initialization to our search.
The third property pertains to the symmetry of the CAID with respect to (w.r.t.) the input alphabet. For the binomial channel ($k=2$), a simple symmetry argument combined with the concavity of the mutual information w.r.t. the input distribution implies that the set of distributions satisfying $f_X((x,1-x))=f_X((1-x,x))$ includes a CAID. The generalized property for the multinomial channel is more involved, and is given as follows. 
\begin{definition}Let $\mathcal{S}_k$ be the set of all bijections from $[k]$ to itself (the symmetric group over $[k]$). An input distribution $f_X(x)$ is said to be invariant under input dimension permutation (IDP) if $f_X(x) = f_{X}(\pi(x))$ for any $x\in\Delta_k$  and any $\pi \in \mathcal{S}_k$. \label{def:invariant}
\end{definition}
\begin{lemma} 
There exists a CAID  that is invariant under IDP.
\label{lemma:symmetry}
\end{lemma}
\full{
\begin{IEEEproof} 
With a slight abuse of notation, let us denote the mutual information induced over the multinomial channel for input $f_X$ by $I(f_X)$, and let ${f}^*_X$ be a CAID. Then, the symmetry of the multinomial channel implies that $ I(f^*_X) = I(f^*_{\pi(X)})$.  Let $f^\#_{X} = \frac{1}{k!} \sum_{\pi\in\mathcal{S}_k} f^*_{\pi(X)}$. The distribution $f^\#_{X}$ is invariant under IDP, and the concavity of $I(f_X)$ w.r.t. $f_X$ along with the aforementioned symmetry then implies that $ I({f}^*_X) \leq I(f^\#_{X})$, so that $f^\#_{X}$ is a CAID. In fact, further analysis shows that there exists a CAID ${f}^*_X(x)$ that is invariant under IDP and also has minimal support-size $m$.  

\end{IEEEproof}
}
The above properties allow us to reduce the search space for a CAID from all possible distributions on $\Delta_{k}$ to the subset of input distributions that are finitely supported with at most $m$ atoms and which are invariant under IDP. Thus, we will search a CAID over distributions supported on the $(k-1)$-dimensional ordered simplex $\Delta^{\geq}_k:= \left\{ x \in \Delta_k \mid x_0 \geq x_1 \geq \cdots \geq x_{k-1} \geq 0 \right\}$,
such that the input distribution $f_{X}$ corresponding to a distribution $\tilde{f}_X$ supported on $\Delta^{\geq}_k$ is given by 
\begin{equation} \label{eq: expanding order simplex to entire simplex}
f_{X} = \frac{1}{k!} \sum_{\pi\in\mathcal{S}_k} \tilde{f}_{\pi(X)}.
\end{equation}

\section{Multidimensional Dynamic Assignment Blahut-Arimoto}
\label{sec:MultinomialDAB}

Our proposed algorithm M-DAB generalizes the DAB algorithm, which finds the CAID for the multinomial channel, rather than for the binomial channel. The main novelty of M-DAB is the handling of the multiple dimensions of the multinomial channel. 
To present M-DAB, we first briefly review algorithms for the binomial channel \cite{komninakis2001capacity}, and specifically the DAB algorithm.

\subsection{Preliminaries: Algorithms for the Binomial Channel}
 In \cite{komninakis2001capacity}, it was suggested to solve the dual problem \eqref{eq: capacity optimization problem} using the ellipsoid method\cite{boyd2004convex}. However, \cite{wesel2018efficient} reported that this method converges slowly, even with well-chosen initial conditions. The DAB algorithm was developed to overcome this. Subsequently, \cite{farsad2020capacities} further improved and utilized it to find a CAID of the particle-intensity channel.

The DAB algorithm can be thought of as a primal-dual alternating optimization algorithm, in which the weights and locations of the CAID \eqref{eq: atomic input distribution} are alternatively updated via the primal and dual problems. Specifically, the dual problem is used to update the locations, and the main idea is that given $P_Y$, adding a point in the location of the maximizer $x_{\max}\in\Delta_k$ tends to reduce the dual objective in \eqref{eq: dual problem alt}. The DAB algorithm for the CAID of $C_{n,k=2}$ is as follows:
\begin{enumerate}[start=0]
  \item Initialize the locations of $f_X$ as the locations of the CAID of $C_{n-1,k=2}$. \label{steps:init}
  \item Run the Blahut-Arimoto algorithm on the current locations to optimize the weights, obtain an input distribution $f_X$, and compute the value of the primal objective \eqref{eq: capacity optimization problem}. Also compute the output distribution $P_Y$. \label{steps:BA}
  \item Using $P_Y$, find the maximizer $x_{\max}\in\Delta_k$ of the KL divergence in \eqref{eq: dual problem alt}. \label{steps:max}
  \item Find the nearest mass point $x_{\mathrm{closest}}$ in $f_X$ to  $x_{\max}$. \label{steps:near}
  \item Determine whether to add a point to the support of $f_X$. \label{steps:add}
  \item Move $x_{\mathrm{closest}}$ in the direction of  $x_{\max}$ and compute the value of the dual objective function \eqref{eq: dual problem alt}.\label{steps:move}
  \item Stop if the primal and dual are $\epsilon$-close. Otherwise Jump to step 1. \label{steps:jump}
\end{enumerate}

\subsection{Overview of the M-DAB Algorithm} 

\begin{algorithm}[t]
  \centering
  \begin{algorithmic}[1]
  \renewcommand{\algorithmicrequire}{\textbf{Input:}}
  \renewcommand{\algorithmicensure}{\textbf{Output:}}
  \REQUIRE $\v{x} = (x^{(1)},x^{(2)},\ldots,x^{(m)}), n, \epsilon$.
  \ENSURE $\v{x} = (x^{(1)},x^{(2)},\ldots,x^{(m')}), \v{p} = (p_1,p_2,\ldots,p_m)$ st. $I(\v{x}, \v{p}) \geq C - \epsilon$. 
  
  \STATE $\v{x} \gets ReduceToOrderedSimplex(\v{x})$ \label{MDAB:line:reduce}
  \WHILE {True}
  
  \STATE $\v{p} \gets BlahutArimoto(Expand(\v{x}))$ \label{MDAB:line:BA}
  \STATE $I \gets I(Expand(\v{x}), \v{p})$ \label{MDAB:line:MI}
  \STATE $D, x_{\mathrm{max}} \gets \max\limits_{x \in \Delta^{\geq}_k} D(P_{Y|X=x}||P_{Y}(Expand(\v{x}), \v{p}))$ \label{MDAB:line:D}
  
  \IF{$D - I \leq \epsilon$} \label{MDAB:line:threshold}
  \RETURN $Expand(\v{x}), \v{p}$.
  \ENDIF
  
  
  \STATE $d_x, x_{\mathrm{closest}} \gets \min\limits_{x}  D(x||x_{\mathrm{max}})$ \label{MDAB:line:closest}
  \STATE $d_v, v_{\mathrm{closest}} \gets \min\limits_{v \in Vertices} D(v||x_{\mathrm{max}})$ \label{MDAB:line:vertix}
  
  \IF {$d_x > d_v$} \label{MDAB:line:require}
  \STATE $\v{x} \gets AddPoint(\v{x}, v_{\mathrm{closest}})$ \label{MDAB:line:add}
  \ELSE 
  \STATE $\v{g} \gets CreateDirectionVector(\v{x}, x_{\mathrm{closest}}, x_{\mathrm{max}})$ \label{MDAB:line:direction}
  \STATE $\delta_{\mathrm{max}} \gets \max\limits_{\delta} I\left(Expand(\v{x}+\delta\v{g})  \right)$ \label{MDAB:line:line-search} 
  \STATE $\v{x} \gets \v{x}+\delta_{\mathrm{max}}\v{g}$ \label{MDAB:line:adjust}
  \ENDIF
  
  \ENDWHILE
  
  \end{algorithmic}
  \caption{M-DAB}
  \label{alg:generalDAB}
\end{algorithm}
The M-DAB algorithm (Algorithm~\ref{alg:generalDAB})  is based on the steps of the DAB algorithm. The CAID optimization for $C_{n,k}$ is initialized with the mass points from the CAID of $C_{n-1,k}$, and for $C_{1,k}$ the vertices of the simplex are used, as suggested by Lemma~\ref{lemma:init}. 
An important operation of M-DAB is \textit{ReduceToOrderedSimplex} (row~\ref{MDAB:line:reduce}), which reduces the search space to the ordered simplex by simply removing all the mass points outside the ordered simplex. Whenever a computation requires the full input distribution (rows~\ref{MDAB:line:BA}, \ref{MDAB:line:MI}, \ref{MDAB:line:D} and \ref{MDAB:line:line-search}), we use the expansion operation \textit{Expand} which creates the full simplex from the ordered simplex by inserting all the permutations as suggested in (\ref{eq: expanding order simplex to entire simplex}). 

At each iteration we use the Blahut-Arimoto algorithm (row~\ref{MDAB:line:BA}) over the current locations, in order to get the corresponding weights, which together with the locations uniquely determine the input distribution $f_X$. The input distribution, together with the channel transition probabilities, are then used to calculate the mutual information (row~\ref{MDAB:line:MI}) and the output distribution $P_Y$. Next, we search for the symbol $x_{\mathrm{max}}$, which maximizes the KL divergence between its output distribution and $P_Y$ (row~\ref{MDAB:line:D}). The maximum divergence $D$ is the value of the dual function, and thus it is an upper bound for the capacity. The algorithm will continue until the mutual information is $\epsilon$ close to the upper bound (row~\ref{MDAB:line:threshold}). 

The next steps involve adjusting our guess based on $x_{\mathrm{max}}$, by checking which mass point is closest to $x_{\mathrm{max}}$ using the KL divergence as a distance measure (row~\ref{MDAB:line:closest}). 
In case that one of the vertices of the ordered simplex is closer (row~\ref{MDAB:line:vertix}), this point is added (row~\ref{MDAB:line:add}), using the \textit{AddPoint} operation, which simply inserts the point to $\v{x}$. 
Otherwise, we need to adjust the closest mass point $x_{closest}$. For simplicity, we represent the set $\v{x}$ as a vector. The operation \textit{CreateDirectionVector} (row~\ref{MDAB:line:direction}) is used to create the direction vector $g$, which is $\v{0}$ for any $x$ that is not $x_{closest}$, and there it equals $x_{\mathrm{max}} - x_{closest}$. Then, the algorithm uses a line search (row~\ref{MDAB:line:line-search}) to find the optimal adjustment step $\delta$, and uses it to adjust the locations (row~\ref{MDAB:line:adjust}). 

\subsection{A Generalization to Higher Input Dimensions}
We next emphasize how the multidimensionality $k>2$ affects the steps made in the DAB algorithm, and what modifications were needed.

\textit{Initialization and Step~\ref{steps:init}:} We first harness  the symmetry assured by Lemma~\ref{lemma:symmetry} to reduce the search space. In the binomial channel case, the DAB algorithm utilizes the apparent symmetry between inputs $x\in[0,1]$ and $1-x\in[0,1]$ and adjusts the locations in pairs. M-DAB further expands this symmetry, to make use of all the permutations, and reduce the search space from $\Delta_k$
to $\Delta^{\geq}_k$ (row~\ref{MDAB:line:reduce}), thus reducing the number of the optimized location parameters. The distribution supported on $\Delta^{\geq}_k$ is symmetrically expanded to an input distribution on $\Delta_k$ as in (\ref{eq: expanding order simplex to entire simplex}), and the latter is used to compute the mutual information, for running the  Blahut-Arimoto algorithm, for computing the output distribution, and for finding the maximizer of the KL divergence  (rows~\ref{MDAB:line:BA}, \ref{MDAB:line:MI} and \ref{MDAB:line:D}).
This process is equivalent to changing the initialization of Step~\ref{steps:init} to use only the CAID mass points supported on $\Delta^{\geq}_k$. We also mention that the dependence of each run of the algorithm on an initialization based on previous runs causes the error to accumulate. Step~\ref{steps:jump} includes a threshold, $\epsilon$, such that the algorithm continues until the mutual information is  $\epsilon$-close to the upper bound  (row~\ref{MDAB:line:threshold}). This threshold affects whether the initialization will be sufficient or not for larger values of $n$.

\textit{Step~\ref{steps:BA}:} This step is similar to the DAB algorithm.

\textit{Step~\ref{steps:max}:} This step is challenging in M-DAB since it requires a more complicated multidimensional maximization, compared to a simple line search for $k=2$. Moreover, the bounds of the maximization problem, i.e., the edges of the ordered simplex, tend to behave as a small scale of the original problem. In order to address these challenges we use known the simplicial homology global optimization (SHGO) algorithm \cite{endres2018simplicial}, with sampling using the Sobol sequence \cite{joe2008constructing}.

\begin{figure}[t]
  \centering  \includegraphics[width=1\linewidth]{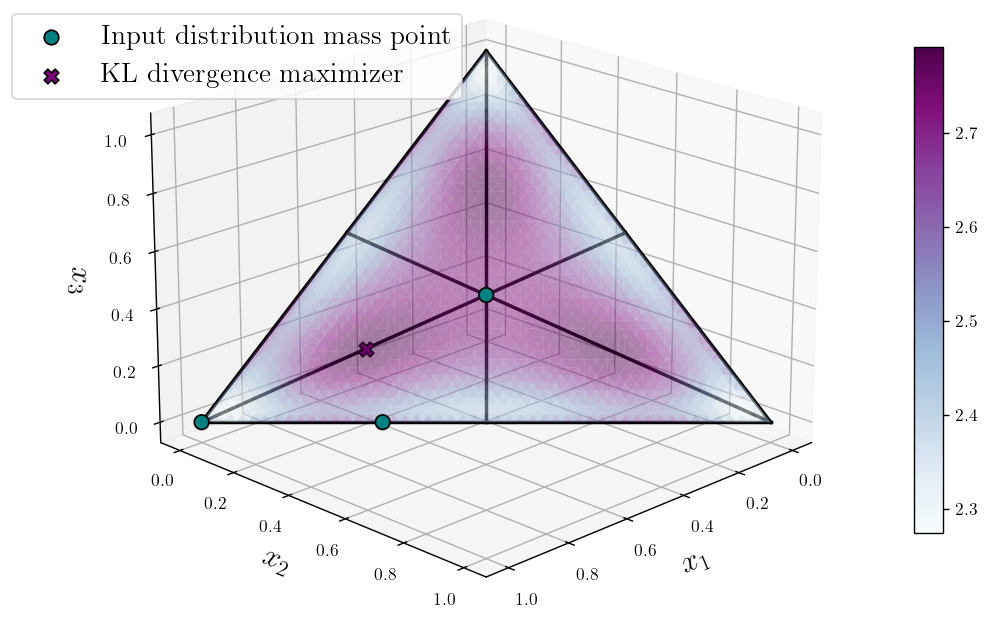}
\caption{The first iteration of the M-DAB algorithm for $C_{n=7, k=3}$. The simplex is an equilateral triangle, and the ordered simplex is a right-angled triangle on the bottom-left. There are 3 mass points in the ordered simplex (blue 'o'). The color bar represents the KL divergence value in each point in the simplex, and the maximizer is marked with purple 'x'.} 
  \label{fig:KL}
  \vspace{-10pt}
\end{figure}

\textit{Step~\ref{steps:near}:} The DAB algorithm finds the closest mass point to the global maximum, and this point is later used in Step~\ref{steps:add} and Step~\ref{steps:move}. This method works well for the binomial case, but in the multinomial case, we experimentally observed iterations in which adjusting the closest point in the direction of the maximum does not increase the value of the mutual information. Moreover, DAB does not adjust the vertices $\{0,1\}$, even if they are the closest ones  \cite{wesel2018efficient}. Later, \cite{farsad2020capacities} suggested to constrain the search of the closest points to the interval bounded by $x_{\mathrm{max}}$ and $0.5$.
We thus conclude that the Euclidean distance is not necessarily the most efficient distance measure, and modify the M-DAB algorithm to use the KL divergence as the distance measure (row~\ref{MDAB:line:closest}). 
The KL divergence is empirically better, and appears to be a natural measure in this scenario, as we would like to find which input is more likely to be interpreted as the global maximum.
Figure~\ref{fig:KL} illustrates such a case. During the first iteration of the algorithm for $C_{n=7, k=3}$ we find the maximum in $(0.616,0.192,0.192)$ while the nearest point using the Euclidean distance is $(0.682,0.318,0)$. This point is on the edge of the simplex, and trying to adjust it is not beneficial, so the algorithm will not converge.   

\textit{Step~\ref{steps:add}:} The M-DAB algorithm decides whether to add a mass point or not. For $k=2$ , DAB adds points either at $x = 0.5$, or by splitting this point. For the multidimensional case, a new mass point may be required in any of the vertices of $\Delta^{\geq}_k$  (rows~\ref{MDAB:line:vertix}, \ref{MDAB:line:require} and \ref{MDAB:line:add}), whereas in the binomial case, the vertices $\{0,1\}$ are always occupied. Moreover, M-DAB algorithm adds a new mass point to the input distribution implicitly whenever a point moves from one of the symmetry axes.

\section{Results}
\label{sec:results}
\begin{figure}[t]
  \centering
  \includegraphics[width=0.965\linewidth]{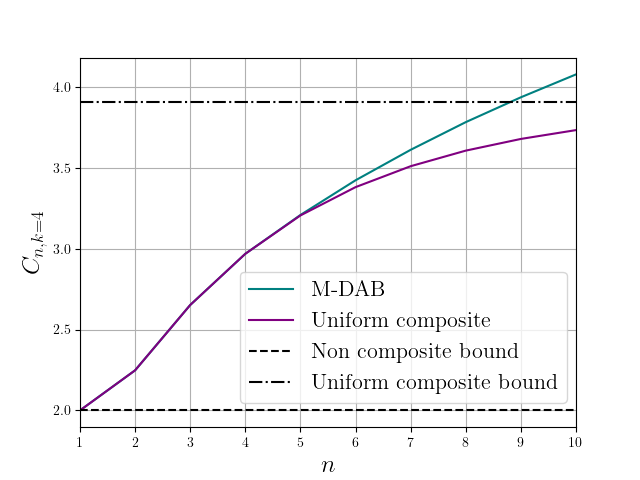}
  \caption{The capacity achieved for $C_{n,k=4}$. In blue, our method M-DAB, and in purple, the method used in \cite{choi2019high}. In black are the upper bounds on any base $4$ and $15$ encoding.}
  \label{fig:CapacityCompare}
  \vspace{-10pt}
\end{figure}
\full{ In this section, we present empirical results of the  M-DAB algorithm. Specifically, for the case $k=4$ suitable for our DNA storage motivation, Table~\ref{tab:locations} in the appendix specifies the mass point locations and weight. These CAIDs can be used in future experiments and systems of composite DNA storage.}
\short{ In this section, we present empirical results of the M-DAB algorithm. Specifically, for the case $k=4$ suitable for our DNA storage motivation, the mass points locations and weights are available in the extended version~\cite{Extended}. These CAIDs can be used in future experiments and systems of composite DNA storage.}
In Figure~\ref{fig:CapacityCompare}, the capacity is achieved by running M-DAB for $k=4$. We compare the mutual information achieved by M-DAB to the naive method of using only the \textit{uniform composite} $(1,0,0,0)$, $(0.5,0.5,0,0)$, $(0.25,0.25,0.25,0.25)$ and $(0.333,0.333,0.333,0)$, proposed in \cite{choi2019high}. 
Considering the critical number of copies where our result surpasses other methods, using composite letters is not beneficial only for $n=1$. Thus, composite letters are strictly better than using ordinary DNA encoding whenever the output strands are redundant, i.e.,  $n\geq2$.  The uniform composite is a CAID for small values of copies, the CAID obtained by M-DAB is better for any $n\geq5$. Furthermore for any $n\geq9$ copies, the capacity of M-DAB surpasses the limit of any base $15$ method ($\log(15) \sim 3.907$).  

\begin{figure}[t]
  \centering
  \includegraphics[width=1\linewidth,keepaspectratio]{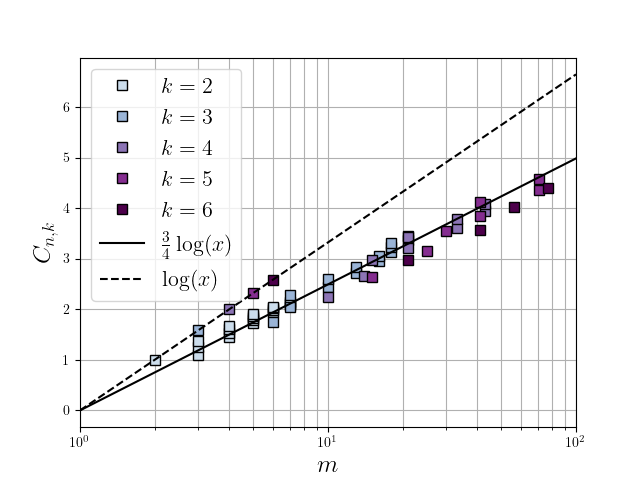}
  \caption{The capacity as function of the number of mass points in the minimal support size CAID. The dimensions are represented in different colors. A scaling law can be observed, where the capacity behaves as a logarithm with a factor of $\frac{3}{4}$.}
  \label{fig:PowerLaw}
  \vspace{-10pt}
\end{figure}
The capacity of the multinomial channel computed by the M-DAB algorithm for finite $n$ allows to compare it with a scaling law recently found in \cite{mattingly2018maximizing} for the binomial and Gaussian channels, which was conjectured to be universal in a subsequent study \cite{abbott2019scaling}. The scaling law claims that the mutual information $I(X;Y)$ for CAID supported on $m$ atoms scales as $\frac{3}{4}\log m$. Our method allows us to plot the capacity and support size for different values of $n$, as a parametric curve (Figure~\ref{fig:PowerLaw}). Doing so we numerically validate the conjecture of \cite{abbott2019scaling}, and testify that M-DAB finds the CAID with the minimum number of mass points. 

\section{Future Research}
\label{sec:conclusion}

We have seen that the dimensionality of the input complicates the design of the M-DAB algorithm compared to the DAB algorithm. While the experimental results demonstrate the effectiveness of M-DAB, theoretical convergence guarantees are lacking. This is challenging since the M-DAB algorithm optimizes the location of just one mass point at each iteration, and so can be viewed as a \textit{coordinate descent} algorithm. The convergence analysis of such algorithms is not always obvious. 
We notice that even for small values of $n$ (such as $n=9$) the CAID requires several dozens of mass points, and as suggested by the scaling law, the support size scales exponentially as a function of the capacity. This might be impractical to implement in DNA storage systems, since any mass point requires a specific mixture of nucleotides, it is desirable to use a minimal number of such mixtures. This raises a natural and important follow-up problem, which is to determine the CAID of the multinomial channel under a constraint on the support size. 

\section*{Acknowledgments}
The research was Funded by the European Union (ERC, DNAStorage, 865630). Views and opinions expressed are however those of the authors only and do not necessarily reflect those of the European Union or the European Research Council Executive Agency. Neither the European Union nor the granting authority can be held responsible for them. The research of N. W. was supported by the Israel Science Foundation (ISF), grant no. 1782/22.

\bibliographystyle{IEEEtran}
\bibliography{refs}

\full{
\newpage
\section*{Appendix A}

\begin{table}[htbp!]	
\centering
\begin{tabular}{|c|c|c|c|c|}

\hline
 $n$ & Mass Points & weights & permutations & total \\
     & Locations   &        &              & number \\
 \hline
 $1$  & $ (1,0,0,0) $& $0.25$ & $ \binom{4}{1} = 4 $ & $4$\\
 \hline
 $2$  & $ (1,0,0,0) $            & $0.172$ & $\binom{4}{1} = 4 $ & $10$ \\ 
      & $(0.5,0.5,0,0) $         & $0.052$ & $\binom{4}{2} = 6$  &\\
 \hline
 $3$  & $(1,0,0,0) $             & $0.130$ & $\binom{4}{1} = 4$ & $14$\\ 
      & $(0.5,0.5,0,0) $         & $0.076$& $\binom{4}{2} = 6$ &\\
      & $(0.333,0.333,0.333,0)$  & $0.006$ & $\binom{4}{3} = 4$ &\\
 \hline
 $4$  & $(1,0,0,0) $             &$0.112$ & $\binom{4}{1} = 4$ & $15$\\ 
      & $(0.5,0.5,0,0) $         &$0.077$ & $\binom{4}{2} = 6$ &\\
      & $(0.25,0.25,0.25,0.25) $ &$0.006$ & $\binom{4}{4} = 1$ &\\
      & $(0.333,0.333,0.333,0)$  &$0.021$ & $\binom{4}{3} = 4$ &\\
 \hline
 $5$  & $(1,0,0,0) $             &$0.097$ & $\binom{4}{1} = 4$ & $21$\\ 
      & $(0.608,0.392,0,0) $     &$0.040$ & $\binom{4}{1, 1} = 12$ &\\
      & $(0.25,0.25,0.25,0.25) $ &$0.004$ & $\binom{4}{4} = 1$ &\\
      & $(0.333,0.333,0.333,0)$  &$0.032$ & $\binom{4}{3} = 4$ &\\
 \hline
 $6$  & $(1,0,0,0) $             &$0.080$ & $\binom{4}{1} = 4$ & $21$\\ 
      & $(0.685,0.315,0,0) $     &$0.043$ & $\binom{4}{1, 1} = 12$ &\\
      & $(0.25,0.25,0.25,0.25) $ &$0.08$ & $\binom{4}{4} = 1$ &\\
      & $(0.333,0.333,0.333,0)$  &$0.039$ & $\binom{4}{3} = 4$ &\\
 \hline
 $7$  & $(1,0,0,0) $             & $0.070$& $\binom{4}{1} = 4$ & $33$\\ 
      & $(0.709,0.291,0,0) $     & $0.043$& $\binom{4}{1, 1} = 12$ &\\
      & $(0.25,0.25,0.25,0.25) $ & $0.012$ & $\binom{4}{4} = 1$ &\\
      & $(0.333,0.333,0.333,0)$  &$0.009$ & $\binom{4}{3} = 4$ &\\
      & $(0.478,0.261,0.261,0)$  &$0.013$ & $\binom{4}{2, 1} = 12$ &\\
 \hline
 $8$  & $(1,0,0,0) $             &$0.063$ & $\binom{4}{1} = 4$ & $29$\\ 
      & $(0.721,0.279,0,0) $     &$0.042$ & $\binom{4}{1, 1} = 12$ &\\
      & $(0.25,0.25,0.25,0.25) $ &$0.016$ & $\binom{4}{4} = 1$ &\\
      & $(0.526,0.237,0.237,0)$  &$0.19$ & $\binom{4}{2, 1} = 12$ &\\
 \hline
 $9$  & $(1,0,0,0) $             & $0.057$ & $\binom{4}{1} = 4$ & $39$\\ 
      & $(0.5,0.5,0,0) $         & $0.010$& $\binom{4}{2} = 6$ &\\ 
      & $(0.25,0.25,0.25,0.25) $ & $0.016$& $\binom{4}{4} = 1$ &\\
      & $(0.552,0.224,0.224,0)$  & $0.021$& $\binom{4}{2, 1} = 12$ &\\
      & $(0.748,0.252,0,0) $     & $0.036$ & $\binom{4}{1, 1} = 12$ &\\
      &$(0.427,0.191,0.191,0.191)$& $0.003 $& $\binom{4}{3} = 4$ &\\
 \hline
 $10$ & $(1,0,0,0) $             & $0.052$& $\binom{4}{1} = 4$ & $38$\\ 
      & $(0.5,0.5,0,0) $         & $ 0.020$& $\binom{4}{2} = 6$ &\\ 
      & $(0.568,0.216,0.216,0)$  & $ 0.022$& $\binom{4}{2, 1} = 12$ &\\
      & $(0.782,0.218,0,0) $     & $ 0.031$& $\binom{4}{1, 1} = 12$ &\\
      &$(0.430,0.190,0.190,0.190)$& $0.009$& $\binom{4}{3} = 4$ &\\
 \hline

\end{tabular}
\caption{Mass Points for $C_{n,k=4}$}
\label{tab:locations}
\end{table}
}

\end{document}